\documentstyle[12pt]{article}

\renewcommand{\theequation}{\arabic{section}.\arabic{equation}\alph{subeq}}
\newcounter{subeq}
\newcommand{\bml}[1]{\mbox{\large\boldmath${#1}$}}

\newcommand{\be}{\begin{equation}}
\newcommand{\ee}{\end{equation}}
\newcommand{\bea}{\begin{eqnarray}}
\newcommand{\eea}{\end{eqnarray}}
\newcommand{\s}{S[\Omega;\Lambda]}
\newcommand{\sdl}{S[\Omega;\Lambda (\delta t)]}

\newcommand{\sfulldlg}{
 \sum_n {1\over n!}\int {d^4 p_1\over (2\pi)^4}
 \cdots {d^4 p_n\over (2\pi)^4}(2\pi)^4\delta^4(p_1+\cdots +p_n)\nonumber\\
  & &\theta (\Lambda (\delta t) -|p_1|)\cdots \theta (\Lambda (\delta t)-|p_n|)
 g_{i_1\cdots i_n}(p_1,\ldots,p_n;\Lambda (\delta t))\nonumber\\
  & & \Omega_{i_1}(-p_1) \cdots \Omega_{i_n}(-p_n) }
\newcommand{\ft}{
 \sum_n {1\over n!}\int {d^4 p_1\over (2\pi)^4}
 \cdots {d^4 p_n\over (2\pi)^4}(2\pi)^4\delta^4(p_1+\cdots +p_n)}
\newcommand{\fields}{\Omega_{i_1}(-p_1) \cdots \Omega_{i_n}(-p_n) }

\begin{document}
\hfill PURD-TH-92-9\\
\vspace{0.5in}
\begin{center}
{\large\bf WILSON RENORMALIZATION GROUP ANALYSIS OF THEORIES
WITH SCALARS AND FERMIONS}
\end{center}
\vspace{0.3in}
\begin{center}
{\bf T.E. Clark, B. Haeri,  S.T. Love}\\
 ~\\
{\it Department of Physics\\
Purdue University\\
West Lafayette, IN 47907-1396}
\end{center}
\vspace{0.5in}
\begin{center}
{\bf Abstract}
\vspace{10pt}
\end{center}

\newcommand{\spone}{0.9}
\newcommand{\sptwo}{1.4}
\newcommand{\singlespace}{\edef\baselinestretch{\spone}\Large\normalsize}
\newcommand{\doublespace}{\edef\baselinestretch{\sptwo}\Large\normalsize}
\singlespace

The continuous block spin (Wilson) renormalization group equation governing
the scale dependence of the action is constructed for theories containing
scalars and
fermions.  A locally approximated form of this equation detailing the structure
of a generalized effective potential is numerically analyzed.  The role of the
irrelevant operators in the nonperturbative renormalization group running is
elucidated and a comparison with the 1-loop perturbative results is drawn.
Focusing on the spontaneously broken phase of a model possessing a discrete
symmetry forbidding an explicit fermion mass term, mass bounds on both the
scalar and fermion degrees of freedom are established.  The effect of the
generalized Yukawa coupling on the scalar mass upper bound is emphasized.

\pagebreak

\doublespace

\section{Introduction}

A basic, generally unstated, tenet used when describing
elementary particle interactions is that the physics at
one distance scale follows uniquely from the dynamics
specified on a more finely grained distance scale.  As such,
it is crucial to understand the short distance dynamics
of the standard model of elementary particle interactions.
Since the non-Abelian gauge interactions are asymptotically
free, their short distance behavior is amenable to a
renormalization group improved perturbative analysis.
On the other hand, the scalar self interactions and the
scalar-fermion Yukawa interactions are characterized by couplings which are
not asymptotically free.  Consequently any conclusions
drawn from a perturbative analysis of their short distance
structure is at best indicative and require confirmation
using nonperturbative techniques.

For instance, a study of the one-loop effective potential
in the self coupled $\lambda \phi^4$ scalar model produces
a Landau singularity$^{[1]}$ which has in turn been used to place a
perturbative triviality bound$^{[2]}$ on the scalar mass.  The
qualitative nature of these conclusions has been largely
substantiated by various nonperturbative studies.$^{[3-4]}$  Here it
is found that a nontrivial self coupled scalar model can
only be consistently defined as a cutoff theory; that is, as
an effective description of the dynamics below some momentum
scale $\Lambda$.  The only consistent $\Lambda \rightarrow \infty$
limit of this model is one characterized by a vanishing
renormalized coupling.  This free field behavior is a reflection
of the Landau singularity already observed in renormalization
group improved perturbation theory.

Within the perturbative framework, given the value of the four
scalar self coupling at the scale $\Lambda,~\lambda(\Lambda)$,
which in general is such that $\lambda(\Lambda) > 1$, then
the coupling at smaller momentum scales follows from the
one-loop renormalization group equation as
\be  \frac{1}{\lambda(m_H)} = \frac{1}{\lambda(\Lambda)}
    + \frac{3}{2\pi^2} {\rm ln}~\frac{\Lambda}{m_H} \, ,
\label{1.1}
\ee
where $m_H$ is the scalar mass.  In general, this equation
is inadequate as it is being applied in a range outside the
domain of validity of perturbation theory.  If however,
$m_H << \Lambda$, then the low energy value $\lambda(m_H)$
is insensitive to the initial value $\lambda(\Lambda)$
due to the renormalization group trajectory focusing effects
of the trivial (Gaussian) infrared fixed point for $\lambda$.
On the other hand, when discussing upper bounds on the
scalar mass, one must consider values of $m_H$ which are
not much smaller than $\Lambda$ itself.  For instance
for $\Lambda \sim 1$ Tev, the upper bound on $m_H$ using perturbation
theory$^{[2]}$ is found to be $m_H \sim 800$ GeV.  Certainly
these quantitative results require substantiation in a
nonperturbative framework.  In fact it turns out that
various such estimates$^{[3-4]}$ have been obtained yielding a
maximum scalar mass of $m_H \sim 650-800$ GeV again for
$\Lambda \sim 1$ TeV.  This rough quantitative agreement of
these nonperturbative predictions with the perturbative
estimate is certainly not a priori anticipated.  In addition to the
fact that the scalar self coupling is sufficiently large
to render the 1-loop perturbative result suspect, it also
follows that such a coupling induces additional irrelevant
operators whose effects are not necessarily insignificant
for scales not too far removed from $\Lambda$.  In
general, the coefficients of these operators are suppressed
by powers of $m_H/\Lambda$.  While for $m_H << \Lambda$
such terms are clearly quite small, this need not
be the case as $m_H$ approaches $\Lambda$.

Of the various nonperturbative studies which have
been performed for these scalar self coupled systems, most
have employed lattice Monte Carlo simulations.$^{[3]}$   A somewhat
different approach$^{[4]}$  to the study of these systems employs the
continuous
Wilson renormalization group.$^{[5-7]}$    In this case, one specifies
the model at the ultraviolet (UV) cutoff scale $\Lambda$ and, by demanding
the cutoff independence of the renormalized Green functions,
one constructs a functional differential equation for the
action at the longer distance scale. This action contains
not only the relevant and marginal operators but also
the irrelevant operators with the coefficient of each operator
fixed in terms of the initial action defined at scale $\Lambda$.
Moreover as the construction is independent of the strength
of the initial coupling, it constitutes a nonperturbative
procedure.  Using a local action approximation$^{[8]}$ to these
exact functional equations, Hasenfratz and Nager$^{[4]}$ extracted
an upper bound for the scalar mass consistent with the other
nonperturbative estimates.  Recently, the sensitivity of
this bound to the choice of action at scale $\Lambda$ has
also been investigated.$^{[9]}$    The resulting maximum scalar
mass was found to be insensitive to the initial action.

The purpose of this paper is to extend the Wilson
renormalization group approach to the construction of
renormalized effective Lagrangians in theories containing
fermions in addition to scalars.  Once again, a consistent
model of this sort requires the introduction of an ultraviolet
cutoff $\Lambda$.  Provided that a huge hierarchy exists
between $\Lambda$ and the scalar and fermion masses (generally
accomplished by a fine tuning of parameters), then the
infrared stable quasi fixed point$^{[10]}$  of the Yukawa and scalar
self couplings again focuses the renormalization group trajectories
to a low energy value basically insensitive to the exact nature
of the short distance couplings.  Moreover the position of
the infrared quasi fixed point can be determined using the
perturbative 1-loop renormalization group equations for these
couplings.  This is precisely the situation which has been
exploited in the top quark condensate models.$^{[11]}$

On the other hand, for scalar and fermion masses not too far
removed from the UV cutoff, some nonperturbative tools
must be employed.  In the next section, we use the continuous Wilson
renormalization group to construct a functional differential
equation for the action functional which can serve as just
such a nonperturbative procedure.  To obtain this equation,
one varies the UV cutoff and demands the resultant action
functional reproduces the same physics (i.e. renormalized
Green functions) on all momentum scales less than the new
(lower) cutoff.  That is, the change in cutoff is compensated
by changing the coupling constants of a complete set of
local operators.  This set includes irrelevant operators in
addition to the relevant and marginal operators.  Thus the
various action functionals satisfying this equation
correspond to points on a particular (Wilson) renormalization
group flow each producing the same physics on momentum
scales less than the UV cutoff used in its definition.

This exact Wilson renormalization group equation is
tantamount to an infinite number of coupled consistency
equations for the various coupling constants of the complete
set of operators.  As such, its analysis and extraction of
physical results from it requires additional approximations.
In section 3, we employ a local action approximation
in terms of which the functional differential equation
reduces to a partial differential equation for a generalized
effective potential in a model containing one scalar and one fermion.
In the limit of small scalar self
coupling and Yukawa coupling at the UV cutoff scale, this
equation reduces to the familiar 1-loop perturbative renormalization
group equations for these couplings.  For larger
couplings at the UV cutoff scale, however, the equation for the generalized
effective
potential requires a numerical evaluation.  We perform such an analysis
under the further assumption that only terms up to bilinear in the fermion
fields are retained in the generalized potential.  The resulting numerical
solutions are then analyzed for various choices of the initial Yukawa and
scalar self couplings and for different initial cutoffs.  In particular, the
role of the induced irrelevant operators in driving the theory toward the
trivial (Gaussian) fixed point is emphasized.  The exact numerical solutions
are also contrasted with the analytic
solutions of the one loop perturbative equations extrapolated well outside
their domain of validity.

By examining the numerical solutions resulting from different values of the
scalar self
coupling and Yukawa coupling chosen at various cutoff
scales and resulting in nontrivial scalar vacuum expectation values, we
obtain in section 4 the trivialty and vacuum stability bounds on both the
scalar and fermion masses.  Our results
indicate that the scalar mass bound increases in the presence of the Yukawa
coupling.  The amount of this increase depends on the strength of the initial
scalar self coupling; larger self couplings yielding smaller percentage
increases.  For the range of couplings we considered, the percent increase in
the scalar mass bound varied from roughly 15-20\% to 6-10\%.  Similar
conclusions have also recently
been reached from lattice Monte Carlo simulations of scalar-Yukawa
theories.$^{[12]}$    In addition, using the values of the mass
upper bounds as a function of the cutoff, we determine the allowed ranges of
scalar and fermion masses for a specific choice of the cutoff.
\newpage

\section{Wilson Renormalization Group Equation}

\setcounter{equation}{0}

Consider a general Euclidean quantum field theory containing the fields
$\Omega_i$.  The subscript $i$ delineates bosonic or fermionic
fields, as well as labels any Lorentz and internal symmetry indices,
e.g. $\Omega_i \in \{\varphi,\psi, \bar\psi,A_\mu,\,...\}$.  The
theory is regulated
so that the contributions to the Schwinger functions (the Euclidean space Green
functions) of the degrees of freedom
with momentum above $\Lambda$ are strongly damped, and it is
renormalized so that the
Schwinger functions are independent of the cutoff $\Lambda$.  Thus the
generating functional for the renormalized Schwinger functions, $Z[J]$,
is cutoff independent: ${d\over d\Lambda}
Z[J]=0$.  In general, $Z[J]$ may be represented by the path integral
\be
Z[J]=\int [d\Omega] e^{-S[\Omega;\Lambda]+\int d^4 x J_i \Omega_i} \, .
\label{2.1}
\ee
Here $S[\Omega;\Lambda]$ is the Euclidean action, assumed known at scale
$\Lambda$ where the coupling constants are specified while
the sources $J_i$ are of complimentary character to the fields $\Omega_i$ and
have support only on momentum scales below $\Lambda$.  More specifically,
the fields depend on the chosen regulation scheme, $\Omega_i=\Omega_i
(p;\Lambda)$, while the action has the form
\bea
S[\Omega;\Lambda]&\equiv &\sum_n {1\over n!}  \int  {d^4 p_1\over (2\pi)^4}
\cdots {d^4 p_n\over (2\pi)^4}\;(2\pi)^4\delta^4(p_1+\cdots +p_n)\nonumber\\
& &g_{i_1\cdots i_n}(p_1,\ldots,p_n;\Lambda)\,
\Omega_{i_1}(-p_1;\Lambda) \cdots \Omega_{i_n}(-p_n;\Lambda) \, ,
\label{2.2}
\eea
with given coupling constants $g(p;\Lambda)$.  The detailed question
of regulator dependence of $S$ will be addressed in future work.  For
simplicity, in the following, we consistently employ a hard cutoff regulator
in momentum space.
Thus the degrees of freedom with momentum above $\Lambda$ are taken
to be zero and we can write $\Omega(p;\Lambda)\equiv \theta
(\Lambda -|p|) \Omega (p)$.  The action then becomes
\newpage
\bea
S[\Omega;\Lambda]&=&\sum_n {1\over n!} \int {d^4 p_1\over (2\pi)^4}
\cdots {d^4 p_n\over (2\pi)^4}\;(2\pi)^4\delta^4(p_1+\cdots +p_n)
\theta (\Lambda -|p_1|)\nonumber\\
& &\cdots \theta (\Lambda -|p_n|)g_{i_1\cdots i_n}(p_1,\ldots,p_n;\Lambda)
\Omega_{i_1}(-p_1) \cdots \Omega_{i_n}(-p_n) \, , \nonumber \\
& &
\label{2.3}
\eea
while the normalized functional measure in Eq.~(\ref{2.1}) contains only those
degrees of freedom with momentum less than $\Lambda$:
$[d\Omega] =\prod_{|p|\leq \Lambda}d\Omega (p)$.

To derive the continuous Wilson renormalization group equation, we allow the
cutoff to vary and consider the response of the theory subject to the
constraint that the Schwinger functions remain unaltered.  Parametrizing the
cutoff as $\Lambda (t)=e^{-t}\Lambda$, the dependence of the action
on $\Lambda (t)$, or equivalently on $t$,  can be determined
by integrating out the degrees of
freedom with momentum lying between the differentially close
cutoff values, $\Lambda (t=0)=\Lambda$
and $\Lambda (\delta t)=e^{-\delta t}\Lambda=\Lambda -\delta t \Lambda$.
Towards this end,
we expand the action $S$ around the degrees of freedom with momentum
below $\Lambda (\delta t)$ retaining terms up to linear in $\delta t$.
This requires retaining only terms no higher than quadratic in the
fields with momentum in the differential shell
$e^{-\delta t} \Lambda \leq |p| \leq \Lambda$ since each
momentum integral over the momentum shell
introduces a differential
factor of $\delta t$ (save for the one momentum integration over the overall
energy-momentum conserving delta function).  This Gaussian integral over the
fields with momentum in the shell can be performed yielding an action
corresponding to the cutoff $\Lambda
(\delta t)$, which can then be used in the path integral for the remaining
lower momentum fields.  So doing we find
\be
Z[J]=\int_{|p|\leq \Lambda (\delta t)} [d\Omega]e^{-S[\Omega;\Lambda
(\delta t)]+\int {d^4 p\over (2\pi)^4}J_i (p) \Omega_i (-p)} \, .
\label{2.4}
\ee
The new action defined at the lower cutoff $\Lambda (\delta t)$,
$S[\Omega;\Lambda (\delta t)]$, is thus obtained as
\newpage
\bea
S[\Omega;\Lambda (\delta t)]&=&  S[\Omega;\Lambda (0)]\! {\bml |}
+{1\over 2}\int_{shell} {d^4 p\over (2\pi)^4}
{\rm str}\ln{\left( {M(p,-p)\over \Lambda^2}\right)}\nonumber \\
 & &-{1\over 2}\int_{shell} {d^4 p\over (2\pi)^4}{d^4 q\over (2\pi)^4}\;
(-)^{F_i} \left. {\delta S[\Omega;\Lambda]\over \delta\Omega_i (p)}\right|
\,M^{-1}_{ij}(p,q)\,
\left. {\delta S[\Omega;\Lambda]\over \delta\Omega_j (q)}\right| \, ,
\nonumber \\
\label{2.5}
\eea
where $F_i$ is the Grassmann number (even for bosons, odd for fermions)
of the field $\Omega_i$ and
the vertical lines appearing on the right hand side indicate that each
such term is to be evaluated setting the
fields having momentum greater than $e^{-\delta t}\Lambda$ to zero.
We have further defined the inverse propagator, $M_{ij}(p,q)$, as the second
order expansion coefficient of the action so that
\be
M_{ij}(p,q)\equiv (-)^{\Delta{ij}} \left. {\delta^2 S[\Omega;\Lambda]\over
\delta \Omega_j (q)\delta \Omega_i (p)}\right| \, ,
\label{2.6}
\ee
where $\Delta_{ij}$ is unity if both the field $\Omega_i$ is an even element
of the
Grassmann algebra and the field $\Omega_j$ is an odd element of the Grassmann
algebra and vanishes otherwise.  More explicitly, $M$ has the form
\be
M=\pmatrix{M_{BB} & M_{BF}\cr
M_{FB} & M_{FF}\cr} \, ,
\label{2.7}
\ee
where $M_{BB}$ ($M_{FF}$) is the submatrix of bosonic character
obtained by differentiating
$S$ with respect to two bosonic (fermionic) fields, while the submatrices
$M_{BF}$ and $M_{FB}$ are of fermionic character and
result from one bosonic and one fermionic field derivative.
The superdeterminant
of $M$ is defined in terms of the supertrace of the above component
matrices
(see the Appendix for a review of manipulations with Grassmann valued matrices)
yielding
\be
{\rm sdet}M={{\rm det}M_{BB}\over {\rm det}N_{FF}} \, ,
\label{2.8}
\ee
with
\be
N_{FF}=M_{FF}-M_{FB}M_{BB}^{-1}M_{BF} \, ,
\label{2.9}
\ee
and hence the result
\be
{\rm str~ln}M={\rm tr~ln}M_{BB} -{\rm tr~ln}N_{FF} \, .
\label{2.10}
\ee
The action $S[\Omega;\Lambda]{\bml |}$ is obtained
from $\s$ by setting all modes with momentum above $\Lambda (\delta t)$ to
zero and thus can be expanded as
\bea
S[\Omega;\Lambda]{\bml |} &=&
 \sum_n {1\over n!}\int {d^4 p_1\over (2\pi)^4}
 \cdots {d^4 p_n\over (2\pi)^4}(2\pi)^4\delta^4(p_1+\cdots +p_n)\nonumber\\
  & &\theta (\Lambda (\delta t) -|p_1|)\cdots \theta (\Lambda (\delta t)-|p_n|)
 g_{i_1\cdots i_n}(p_1,\ldots,p_n;\Lambda)\nonumber\\
  & & \Omega_{i_1}(-p_1) \cdots \Omega_{i_n}(-p_n) \, .
\label{2.11}
\eea
(Note that the coupling constants are still those originally defined
at $\Lambda$.)
The new action appearing on the left hand side of Eq.~(\ref{2.5}), on the
other hand, involves the lower momentum
fields as well as the coupling constants defined at the lower cutoff and
can be written as
$\Lambda (\delta t)$
\bea
S[\Omega;\Lambda (\delta t)] &=&\sfulldlg \, .
\label{2.12}
\eea
Their difference just results in the variation of the coupling constants
when the cutoff is lowered so that
\bea
\sdl &-& \s {\bml |} =\ft \nonumber\\
& &\theta(e^{-\delta t}\Lambda -|p_1|)
\cdots \theta(e^{-\delta t}\Lambda -|p_n|)\left[g(p_1,\ldots,p_n;
e^{-\delta t}\Lambda)\right. \nonumber\\
 & &\left. \qquad\qquad - g(p_1,\ldots,p_n;\Lambda) \right]
\fields\nonumber\\
&=&-\delta t \ft \nonumber\\
& &\qquad\theta(\Lambda -|p_1|)
\cdots \theta(\Lambda -|p_n|)\nonumber\\
& &\qquad\quad\left[\Lambda{\partial\over \partial\Lambda}
g(p_1,\ldots,p_n;\Lambda)\right]\fields \, , \nonumber \\
& &
\label{2.13}
\eea
where the overall factor of $\delta t$ allows $t$ to be set to zero
inside the integrand in the last line on the right hand side (we
consistently retain only terms through $\delta t$).

In order to obtain a differential equation for the action functional
$S$ itself, the $\Lambda-$derivative must be brought past the regulated
fields (i.e. the $\theta$-functions).  This can be achieved by rescaling
all the momentum vectors to unit vectors
\be
p^\mu =\Lambda \hat p^\mu
\label{2.14}
\ee
so that $\theta (\Lambda -|p|)=\theta (1-|\hat p|)$ and by rescaling all
the fields to dimensionless ones as
\be
\Omega_i (p) =C^{-1}_{\Omega_i}(\Lambda)\hat\Omega_i (\hat p) \, ,
\label{2.15}
\ee
(i.e. for the regulated fields,
$\Omega_i (p;\Lambda) =C^{-1}_{\Omega_i}(\Lambda)\hat\Omega_i (\hat p;1)$).
It follows that the coupling constant variation now includes a
momentum dependence since
\be
\Lambda {\partial\over\partial \Lambda} g(p;\Lambda) =\Lambda
{d\over d\Lambda} g(\Lambda \hat p;\Lambda)-\sum_{i=1}^{n}
\hat p^\mu_i{\partial\over \partial \hat p^\mu_i}g(\Lambda\hat p;\Lambda) \, .
\label{2.16}
\ee
Substituting these variable changes into Eq.~(\ref{2.13}), and recalling
that $\Lambda(t) = e^{-t} \Lambda$ so that a differential logarithmic
change in $\Lambda$ corresponds to a negative differential change in $t$,
we obtain
\newpage
\bea
\Lambda {d\over d\Lambda} S [\Omega;\Lambda(\delta t)]& \equiv &
- {d\over dt}\hat S [\hat\Omega;t] \nonumber\\
& =& {-1\over \delta t}
\left\{  \sdl  -\s {\bml |}\right\} \nonumber\\
& &\qquad -\left[ 4-\sum_{\Omega_i}\int {d^4\hat p\over (2\pi)^4}\hat\Omega_i
(-\hat p)\right.\nonumber\\
 & &\left.\qquad\qquad \left(d_{\Omega_i}-\gamma_{\Omega_i} -\hat p^\mu
{\partial^\prime\over \partial\hat p^\mu}\right){\delta\over
\delta\hat\Omega_i (\hat p)}\right] \hat S[\hat\Omega;\Lambda] \, ,\nonumber\\
& &
\label{2.17}
\eea
where $\hat S$ is the same action as $S$, but expressed in terms of the
rescaled quantities as
\bea
\hat S[\hat\Omega;t] &=& S[\Omega;\Lambda (t)]\nonumber\\
&=&\sum_n {1\over n!}\int {d^4 \hat p_1\over (2\pi)^4}\cdots
{d^4 \hat p_n\over (2\pi)^4}\;(2\pi)^4\delta^4(\hat p_1 +\cdots +
\hat p_n)\nonumber\\
& &\qquad\theta(1-|\hat p_1|)\cdots\theta(1-|\hat p_n|)\nonumber\\
& &\qquad\quad \hat g_{i_1\cdots i_n}(\hat p_1,\ldots,\hat p_n;\Lambda)
\hat\Omega_{i_1}(-\hat p_1)\cdots \hat\Omega_{i_n}(-\hat p_n) \, ,
\label{2.18}
\eea
with the dimensionless coupling constants defined according to
\bea
\hat g_{i_1\cdots i_n}(\hat p_1,\ldots,\hat p_n;\Lambda)&\equiv &
\left[ \Lambda^{4n-4}\prod_{j=1}^n C^{-1}_{\Omega_{i_j}}(\Lambda)
\right]
g_{i_1\cdots i_n}(\Lambda\hat p_1,\ldots,\Lambda\hat p_n;\Lambda) \, .
\nonumber\\
 & &
\label{2.19}
\eea
In addition, the anomalous dimension, $\gamma_{\Omega_i}$, of each field is
defined by
\be
\Lambda {d\over d\Lambda}\ln{C_{\Omega_i}}(\Lambda)=4-d_{\Omega_i}
+\gamma_{\Omega_i} \, ,
\label{2.20}
\ee
with $d_{\Omega_i}$ the naive field dimension; e.g. $d_{\Omega_i}=1$
for bosons, and $d_{\Omega_i}=3/2$ for fermions.  The prime
on the momentum derivative in Eq.~(\ref{2.17})
indicates that it does not act on the overall energy-momentum conservation
Dirac delta function or $\theta$-functions but only on the
dimensionless couplings in the expression for $\hat S$.

Finally, substituting Eq.~(\ref{2.5}) with $M$ expressed in terms of the
rescaled
momenta and fields into Eq.~(\ref{2.17}),
we obtain the continuous Wilson renormalization group equation for the action:
\bea
{d\over dt} S[\Omega;t]
&=&{1\over 2}{1\over \delta t}\int_{e^{-\delta t}\leq |p|\leq 1}
{d^4 p\over (2\pi)^4}{\rm str~ln}M(p,-p)\nonumber\\
 & &- {1\over 2}{1\over \delta t}\int_{e^{-\delta t}\leq |p|,|q|\leq 1}
{d^4 p\over (2\pi)^4}{d^4 q\over (2\pi)^4} (-)^{F_i}
{\delta S[\Omega;\Lambda]\over
\delta \Omega_i(p)}\nonumber\\
 & &\quad\qquad\qquad\qquad\qquad\qquad\qquad M^{-1}_{ij}(p,q)\, {\delta
S[\Omega;\Lambda]\over \delta \Omega_j(q)}\nonumber\\
& &+\left[4-\sum_{\Omega_i}\int_{|p|\leq 1} {d^4 p\over (2\pi)^4}
\Omega_i (-p)\right. \nonumber\\
 & &\left. \qquad\qquad \left(d_{\Omega_i}-\gamma_{\Omega_i}
+p^\mu {\partial^\prime\over
\partial p^\mu} \right){\delta\over \delta\Omega_i (p)}\right]
S[\Omega;\Lambda] \, .\nonumber\\
 & &
\label{2.21}
\eea
Here and in what follows we drop the carat notation on all quantities
and work only with dimensionless fields and unit momentum vectors.
\newpage

\section{The Local Action Approximation}

\setcounter{equation}{0}

The Wilson renormalization group equation (\ref{2.21}) is an exact
equation determining the action at scale $\Lambda
(t)$ given its initial short-distance form at $\Lambda
=\Lambda (0)$.  Unfortunately its complexity is such that it
can only be solved approximately at the present time.  Towards this end, we
ignore all interactions involving derivative couplings and furthermore
set the wavefunction renormalizations to unity.  This is clearly a drastic
approximation but does not leave the equation completely devoid of content.
Actually, in the small coupling constant regime, this approximation
reproduces the one-loop perturbative renormalization group equations with
anomalous dimensions neglected.
The locally approximated action, $S[\Omega;\Lambda]$, takes the
general form
\bea
S[\Omega;\Lambda]&=&\int d^4 x \frac{1}{2} \Omega_{i_1} (x)
K_{i_1i_2}(\partial_\mu)\Omega_{i_2} (x)\nonumber\\
 & &\qquad\qquad + \sum_n \frac{1}{n!} v_{i_1\cdots i_n}^{(n)}
 \int d^4 x \Omega_{i_1} (x)\cdots \Omega_{i_n} (x)
\nonumber\\
 &=& \frac{1}{2}\int {d^4 p_1\over (2\pi)^4} {d^4 p_2\over (2\pi)^4}
(2\pi)^4\delta^4 (p_1+p_2)\Omega_{i_1}(-p_1) K_{i_1i_2}(p_1)
\Omega_{i_2}(-p_2) \nonumber\\
 & &\qquad\qquad +\sum_n \frac{1}{n!}v^{(n)}_{i_1\cdots i_n}
 \int {d^4 p_1\over (2\pi)^4}
\cdots {d^4 p_n\over (2\pi)^4}(2\pi)^4\delta^4(p_1+\cdots +p_n)
\nonumber\\
 & &\qquad\qquad\qquad\qquad\qquad\Omega_{i_1}(-p_1) \cdots \Omega_{i_n}
(-p_n) \, ,
\label{3.1}
\eea
where the coefficients $v^{(n)}_{i_1\cdots i_n}$ are momentum
independent and it is understood that the Euclidean momentum integrals are
over unit vectors.

When expanding the fields about space-time independent values,
$\Omega_i (x)=\Omega_i$, (or in momentum space
$\Omega_i (p)=(2\pi)^4\delta^4 (p)\Omega_i$), it is convenient
to introduce the \lq\lq generalized potential" $U(\Omega;\Lambda)$ as
\be
U(\Omega;\Lambda)\equiv \sum_n \frac{1}{n!}v^{(n)}_{i_1\cdots
i_n}\Omega_{i_1}\cdots
\Omega_{i_n} \, ;
\label{3.2}
\ee
so that, when evaluated at constant fields, and using $K_{ij}(0)=0$,
the action simply takes the form
\be
S[\Omega={\rm constant};\Lambda]=(2\pi)^4
\delta^4(0)U(\Omega;\Lambda) \, .
\label{3.3}
\ee
Thus for constant fields, the Wilson
renormalization group equation (\ref{2.21}) in this approximation reduces to a
differential equation for this generalized potential given by
\bea
{\partial\over \partial t} U(\Omega,t)&=&\frac{1}{2}\frac{1}{\delta t}
\int_{1-\delta t\leq|p|\leq 1}{d^4 p\over (2\pi)^4} {\rm str~ln}
\left[ K_{ij}(p) +{\partial^2\over \partial\Omega_j\partial\Omega_i}U(\Omega,
t)\right]\nonumber\\
 & &\qquad\qquad +\left[4-\sum_{\Omega_i}d_{\Omega_i}\Omega_i {\partial
\over \partial \Omega_i}\right]U(\Omega,t) \, .
\label{3.4}
\eea
In obtaining this result, we have used the fact
that the \lq\lq tree" graph contributions to Eq.~(\ref{2.21})
(the penultimate term on the right hand side) vanish in
the constant field expansion since it sets the \lq\lq propagator",
$M^{-1}$, momentum to zero,
which is far from the momentum shell $1-\delta t\leq |p|\leq 1$.  In
addition, the approximation of neglecting
the anomalous dimensions, $\gamma_{\Omega_i}=0$, is consistent with the
assumption of no wavefunction renormalization.

In order to proceed further, a particular choice of fields and action
must be made.  For simplicity, we will study the field theory of
one real scalar field $\varphi$ and one Dirac spinor fermion field
$\psi$ and its conjugate $\bar\psi$.  The local approximation action
then has the form
\bea
S[\varphi,\psi,\bar\psi,t]&=&-\frac{1}{2}\int d^4 x \varphi \partial^2
\varphi + \int d^4 x \bar\psi \gamma \cdot \partial \psi \nonumber\\
 & &\qquad\qquad +\sum_{l,m}\frac{1}{l!m!}v^{(l,m)}(t)\int d^4 x \phi^l
 (\bar\psi\psi)^m \, ,
\label{3.5}
\eea
where for the possible fermion interaction terms we have assumed only
products involving scalar bilinears enter.  Thus only functions of
$\bar\psi\psi$ occur
in $U$.  If the discrete $\gamma_5$-symmetry,
\bea
\varphi^\prime &=&-\varphi\nonumber\\
\psi^\prime &=&\gamma_5 \psi\nonumber\\
\bar\psi^\prime &=&-\bar\psi \gamma_5 \, ,
\label{3.6}
\eea
is also imposed, the interaction terms involve monomials with $l+m$ equal
an even interger and so explicit fermion mass terms are forbidden.
Hence the generalized potential is simply
\be
U(\varphi,\psi,\bar\psi,t)= \sum_{l,m}\frac{1}{l!m!}
v^{(l,m)}(t)\int d^4 x \phi^l (\bar\psi\psi)^m \, ,
\label{3.7}
\ee
and the kinetic energy matrix in $\{\varphi,\psi,\bar\psi\}$ space is
\be
K(p)=\pmatrix{p^2&0&0\cr
0&0&-i\gamma^T \cdot p\cr
0&-i\gamma \cdot p &0\cr} \, .
\label{3.8}
\ee
Here the Hermitian Euclidean space $\gamma$-matrices are defined to satisfy
\linebreak
$\{\gamma_\mu , \gamma_\nu\} = 2 \delta_{\mu\nu}$, with $\gamma_5 = \gamma_1
\gamma_2 \gamma_3 \gamma_4$.

Inserting these expressions in the renormalization group equation (\ref{3.4})
and performing some straightforward, but lengthy, algebra we find
the locally approximated form of the continuous Wilson renormalization group
equation for the generalized potential given by
\bea
{\partial\over \partial t}U(\varphi,\sigma;t)&=&4U(\varphi,\sigma;t)
-\varphi U_\varphi(\varphi,\sigma;t)-3\sigma U_\sigma(\varphi,\sigma;t)
\nonumber\\
 &+&\frac{1}{16\pi^2}\ln{\left( 1+U_{\varphi\varphi}(\varphi,\sigma;t)
 \right)}-\frac{1}{4\pi^2}\ln{\left(1+U^2_\sigma (\varphi,\sigma;t)
 \right)}\nonumber\\
 &+&\frac{1}{16\pi^2}\ln{\left(1+ \Sigma (\varphi,\sigma ;t) \right)} \, ,
\label{3.9}
\eea
where we have introduced the $\sigma$ field as $\sigma\equiv \bar\psi\psi$
and have defined the combination
\be
\Sigma \equiv 2\frac{\sigma U_\sigma}{1 + U_\sigma^2}
\left[U_{\sigma\sigma}-U_{\sigma\varphi}\left( {1\over
1+U_{\varphi\varphi}}\right)U_{\varphi\sigma}\right] \, .
\label{3.10}
\ee
In addition, we have denoted the derivatives of the potential with respect to
the field variables by subscripts on the generalized
potential, so that, for example, $U_{\varphi\sigma}={\partial^2\over
\partial\varphi\partial\sigma}U$.
Finally, we shall assume that the model at the UV cutoff scale is simply
given by the canonical kinetic matrix $K$ of Eq.~(\ref{3.8}) and the initial
generalized potential function
\be
U(\varphi,\sigma;t=0)=\frac{1}{2}\mu^2 (0) \varphi^2 +\frac{1}{4}\lambda (0)
\varphi^4 +g(0)\varphi\sigma \, ,
\label{3.11}
\ee
which is invariant under the discrete $\gamma_5$-symmetry of Eq.~(\ref{3.6}).
Note
again that this symmetry forbids the appearance of an explicit fermion mass
term or equivalently a term linear in $\sigma$.

We seek solutions for the generalized potential in which this discrete
$\gamma_5$-symmetry is spontaneously broken.  This is achieved by choosing
the parameters of the model at the cutoff scale $\Lambda$ such that the vacuum
energy is minimized by the scalar field $\varphi$ acquiring a nonzero vacuum
expectation value.  At the same time, we assume that no fermion bound states
or condensates form.  As such, the vacuum energy is identified with the
potential
\be
V(\varphi;t) = U(\varphi, \sigma = 0;t) \, .
\label{3.12}
\ee
That is, the vacuum is characterized by $\varphi \not= 0 $ but with
$\sigma = 0$.

In order to proceed with the numerical solution of the partial differential
equation, we make the further truncation of retaining terms in the generalized
\newpage
\noindent
potential only through those linear in $\sigma$ (i.e. $\bar\psi \psi$) so that
\footnote{Since $\bar\psi \psi \equiv \sigma$ is a nontrivial even element of
the Grassmann algebra and since no fermion condensate forms, then the general
expansion for $U$ terminates after terms of order $(\bar\psi \psi)^4 \equiv
\sigma^4$.}

\be
U(\varphi, \sigma;t) = V(\varphi;t) + \sigma G(\varphi;t) \, .
\label{3.13}
\ee
Here $V(\varphi;t)$ is the effective potential function defined in
Eq.~(\ref{3.12}),
while $G(\varphi;t)$ is a generalized Yukawa coupling.  This approximation,
made for technical reasons only, neglects those irrelevant operators
containing four or more fermion fields.  Introducing the function
\be
F(\varphi;t) = V_{\varphi} (\varphi;t)
\label{3.14}
\ee
and recalling that $\sigma$ vanishes in the vacuum, then the partial
differential equation for $U$ reduces to the two coupled partial differential
equations for $F$ and $G$ given by
\bea
\frac{\partial F}{\partial t} &=& 3F - \varphi F_{\varphi} + \frac{1}{16 \pi^2}
\frac{F_{\varphi \varphi}}{1 + F_\varphi} - \frac{1}{2 \pi^2} \frac{G
G_\varphi}
{1 + G^2}
\nonumber\\
\frac{\partial G}{\partial t} &=& G - \varphi G_\varphi + \frac{1}{16 \pi^2}
\frac{G_{\varphi \varphi}}{1 + F_\varphi} - \frac{1}{8 \pi^2} \frac
{G G_{\varphi}^{2}}{(1 + G^2)(1 + F_\varphi)} \, .
\label{3.15}
\eea
{}From the functions $F$ and $G$, the running scalar self coupling, $\lambda
(t)$, and running Yukawa coupling, $g(t)$, are defined as
\bea
\lambda (t) &=& \frac{1}{6} F_{\varphi \varphi \varphi} (0;t)
\nonumber \\
g(t) &=& G_\varphi (0;t)
\label{3.16}
\eea
and are plotted in Figures 1-2 for various choices of their initial values at
the UV cutoff $\Lambda~(t = 0).$

\begin{figure}
\vspace{3.0in}
\caption{Scalar self coupling $\lambda(t)$ as a function of $t$ for
various initial Yukawa couplings and cutoff to scalar vacuum expectation
values.}
\vspace{3.0in}
\caption{Yukawa coupling $g(t)$ as a function of $t$ for
various initial scalar self couplings and cutoff to scalar vacuum expectation
values.}
\end{figure}

Using Eqs.~ (\ref{3.11}), (\ref{3.15}) and (\ref{3.16}), it is readily seen
that $g(t)$ decreases as $t$ increases.  On the
other hand, $\lambda (t)$ can either initially increase or decrease depending
on whether $g(0)$ is greater than or less than the value, $g_c (0)$, which is a
function of $\lambda (0)$ and $\mu^2 (0)$ and produces a zero initial
$\lambda$ slope:
$\frac{d}{dt} \lambda (0) = 0$.  This value is readily computed
as
\be
g^2_c (0) = \frac{3}{2} \frac{\lambda (0)}{1 + \mu^2 (0)} \, .
\label{3.17}
\ee
For $g (0) >(<) g_c (0)$, $\lambda (t)$ initially increases
(decreases) as is readily observed in Fig. 1.  In either case, the effect of
the initially large (nonperturbative) $\lambda (0)$ is such as to induce
non-negligible higher dimensional (irrelevant) operators in the action and
consequently to drive the generalized effective potential away from the surface
spanned by the relevant and marginal operators only as was initially the case
(cf. Eq.~(\ref{3.11})).  These induced operators play an important role in
the dynamics for small to moderate $t$ values.

As a measure of their importance, we display in Fig. 3 the evolution of the
dimension 6 operator coupling
\be
\chi (t) \equiv \frac{1}{5!} e^{-2t} F_{\varphi \varphi \varphi \varphi
\varphi} (0;t) \, .
\label{3.18}
\ee
Here we have included the explicit scale factor of $e^{-2t} \sim \frac{1}
{\Lambda^2}$ accompanying the dimension 6 operator in the definition of the
coupling rather than in the operator itself.  As seen in Fig. 3, this induced
coupling rises rapidly to a peak value before beginning to tail off as $t$ is
increased.  For small to moderate t values, this coupling can be quite large
and is instrumental in helping to drive the system towards the Gaussian fixed
point at the origin of coupling constant space.  That is, the effect of this
and the other higher dimensional operators is to dampen the $\lambda (t)$
coupling so that eventually the theory evolves into a region where it can
accurately be described using a perburbative expansion for the generalized
effective potential.

\begin{figure}
\vspace{3.0in}
\caption{The $t$-evolution of the coupling $\chi(t)$ of a dimension 6
operator for various ratios of the cutoff to scalar vacuum
expectation value when the initial couplings are $\lambda(0)=5$ and $g(0)=1$.}
\end{figure}

It is clear from Fig. 1 that there is a range of $t$ values greater than a
value \footnote{A more precise definition of $t^*$ will be provided in
section 4.}
$t^*$ for which $\lambda (t)$ can be well approximated by a linear fuction of
$t$.  Over this range, one can therefore write
\be
U(\varphi, \sigma; t) = U(\varphi, \sigma; t^*) + (t - t^*) \frac{\partial}
{\partial t} U(\varphi, \sigma; t^*) \, .
\label{3.19}
\ee
Moreover, the coefficient of $(t - t^*)$ can be directly gleaned using the
locally approximated Wilson renormalization group equation (\ref{3.9}) for the
generalized effective potential.  So doing, one readily computes for this range
of $t$ values that
\bea
\lambda (t) &=&\lambda (t^*) +(t-t^*){1\over 96\pi^2}\left[
\frac{F_{\varphi\varphi\varphi\varphi\varphi}(0;t^*)}{1+F_\varphi (0;t^*)}-
\frac{108\lambda^2(t^*)}{(1+F_\varphi (0;t^*))^2}\right]\nonumber\\
 & &+(t-t^*){1\over 6\pi^2}\left[3g^4(t^*)
-2g(t^*)G_{\varphi\varphi\varphi}(0;t^*)\right] \, .
\label{3.20}
\eea

This same result for $\lambda (t)$ can also be secured by computing the
generalized effective potential for this range of $t>t^*$ values using the
1-loop perturbative result and then retaining terms linear in $t - t^*$.
The 1-loop contribution to the generalized effective potential is given by
\bea
U (\varphi, \sigma; t) &=& U (\varphi, \sigma; t^*)\nonumber\\
& &+{1\over 2}\int_{e^{-(t-t^*)}\leq |p|\leq 1} {d^4p\over (2\pi)^4}
{\rm str~ln}\left[ K_{ij} +{\partial^2\over \partial\Omega_j\partial\Omega_i}
U(\varphi, \sigma; t^*)\right] \, . \nonumber \\
& &
\label{3.21}
\eea
Once again if we retain only the terms through linear in $\sigma$ and
decompose $U(\varphi, \sigma; t)$ as in Eq.~(\ref{3.13}), we isolate
\bea
V(\varphi; t)&=&V(\varphi; t^*) +
{1\over 2}\int_{e^{-(t-t^*)}\leq |p|\leq 1} {d^4p\over (2\pi)^4}
{\rm ln~}\left[ p^2 +V_{\varphi\varphi}(\varphi;t^*)\right]\nonumber\\
 & &-2\int_{e^{-(t-t^*)}\leq |p|\leq 1} {d^4p\over (2\pi)^4}{\rm ln~}\left[p^2
+
G^2(\varphi;t^*)\right]
\nonumber\\
\label{3.22}
\eea
\bea
G(\varphi; t)&=&G(\varphi; t^*)+{1\over 2}G_{\varphi\varphi}(\varphi;t^*)
\int_{e^{-(t-t^*)}\leq |p|\leq 1} {d^4p\over (2\pi)^4}{1\over p^2 +
V_{\varphi\varphi}(\varphi;t^*)}\nonumber\\
 & &-G(\varphi; t^*)G_\varphi^2 (\varphi;t^*)\nonumber\\
 & &\quad \int_{e^{-(t-t^*)}\leq |p|\leq 1}
{d^4p\over (2\pi)^4} {1\over (p^2+G^2(\varphi;t^*))
(p^2+V_{\varphi\varphi}(\varphi;t^*))}.
\nonumber\\
\label{3.23}
\eea
The momentum integrations are standard and can be staightforwardly performed.
Retaining only terms linear in $(t-t^*)$, which constitutes a reasonable
approximation for the $t$ range under consideration ($0<<t-t^*<1$), the
expression for $\lambda (t)$ is found to be identical with that of
Eq.~(\ref{3.20}).
It is important to note that the 1-loop perturbative result includes the
effects of the higher dimensional operators.  Their presence is crucial for
allowing a smooth joining with the numerically generated solution of the
locally approximated Wilson renormalization group equation.

Of course, as $t$ continues to increase to even larger values, the effects of
all the higher dimensional operators will eventually decouple and can be
completely ignored. (They are truly
irrelevant!)  In addition,
the marginal operators will have couplings well in the perturbative region.
The onset of this behavior is more rapidly achieved the larger the value of the
initial cutoff $\Lambda$.  For this range of $t$ values, the evolution of the
model can be accurately parametrized by summing the leading logorithms of the
1-
loop graphs thus obtaining the 1-loop perturbative renormalization group
running for $\lambda (t)$.

To perform the perturbative analysis, we reconsider the partial differential
equations (\ref{3.15}) this time for small values of the initial couplings.
Expanding $F$ and $G$ as
\bea
F(\varphi;t) &=& \mu^2 (t) + \lambda (t) \varphi^3 + ...
\nonumber \\
G(\varphi;t) &=& g(t) \varphi + ...
\label{3.24}
\eea
and retaining only the lowest order terms, the differential equations
reduce to
\bea
16 \pi^2 \frac{d\lambda (t)}{dt} &=& -18 \lambda^2 (t) + 8 g^4 (t)
\nonumber \\
16 \pi^2 \frac{dg(t)}{dt} &=& - 2 g^3 (t)
\label{3.25}
\eea
which are precisely the one-loop perturbative renormalization group
equations in the approximation in which the anomalous dimensions are neglected.
In obtaining these equations, we have assumed that the parameter $\mu^2 (0)$
has been tuned to be close to its critical value, $\mu^2_{crit} = - \lambda (0)
(v^2/\Lambda^{2})$, which is proportional to $\lambda(0)$ and hence small.
Here we denoted
the $\varphi$ vacuum expectation value by $v/\Lambda$.  The
analytic solution to these equations is readily secured as
\bea
\frac{1}{g^2 (t)} &=& \frac{1}{g^2 (0)} + \frac{1}{4 \pi^2} t
\nonumber \\
\lambda (t) &=& 4 g^2(t) \left[\frac{a(t) + 1}{\sqrt{37}(a(t) - 1) - (a(t) +
1)}
\right] \, ,
\label{3.26}
\eea
where
\be
a(t) = \left( \frac{g^2 (0)}{g^2 (t)} \right)^{\sqrt{37}} \left[\frac{\sqrt{37}
+ 1 + 4 \frac{g^2 (0)}{\lambda (0)}}{\sqrt{37} - 1 - 4 \frac{g^2 (0)}{\lambda
(0)}}
\right] \, .
\label{3.27}
\ee
Extrapolating these solutions well outside their domain of validity, we
display these perturbative results for the running couplings as the dashed
lines
in Figs. 4-5 using the same initial inputs as used in the numerical solution to
the nonperturbative locally approximated Wilson renormalization group
equation.

As can be seen in the $g(0)=5,~\lambda(0)=1$ case, the perturbative
renormalization group running of $\lambda (t)$ and $g(t)$  exhibits
the same qualitative features as the nonperturbative evolution, but
are quantitatively quite different.  The large Yukawa coupling again
initially causes $\lambda (t)$ to increase.  This time, the growth is quite
dramatic, as depicted by the dashed line curve in Fig. 4, since the higher
dimensional operators or even higher loop self coupling effects which provide
the additional damping are now absent.  This rise in $\lambda(t)$ continues
until $g(t)$ has decreased and
$\lambda (t)$ has increased sufficiently so that $g^2 (t)={3\over 2}\lambda
(t)$.  At this point the slope of the curve is zero.  Subsequently, the pure
scalar self coupling term in Eq.~(\ref{3.25}) drives $\lambda (t)$ slowly
towards zero.

\begin{figure}
\vspace{2.5in}
\caption{Comparison of the numerically integrated locally approximated
Wilson (solid line) and 1-loop perturbative (dashed line)
renormalization group equation solutions for the running of the scalar self
coupling for various initial Yukawa couplings and cutoff to
scalar vacuum expectation value ratios.}
\end{figure}

For the case of a smaller initial Yukawa coupling constant and larger
initial scalar self coupling, $g(0)=1,~\lambda (0)=5$,
$g(t)$ is seen in Fig. 5 to be completely described by its perturbative
renormalization
group evolution, while Fig. 4 shows $\lambda (t)$ monotonically decreasing
from its
initial value since $g^2 (0)\leq {15\over 2}{1\over 1+\mu^2 (0)}$.  This is
again in crude qualitative agreement with the exact solution, although the
decrease in $\lambda (t)$ in the exact solution is far more rapid due to the
effects of the induced higher dimensional operators (cf. Fig. 1).  On the other
hand, for
the larger cutoff, $\Lambda =38.6 v$, the effects of the
higher dimensional operators are far less significant and the $t$-evolution of
$\lambda (t)$ runs essentially as in 1-loop renormalization group improved
perturbation theory.

\begin{figure}
\vspace{2.5in}
\caption{Comparison of the numerically integrated locally approximated
Wilson (solid line) and 1-loop perturbative (dashed line)
renormalization group equation solutions for the running of the Yukawa
coupling for various initial scalar self couplings and
cutoff to scalar vacuum expectation value ratios.}
\end{figure}

\newpage

\section {Scalar and Fermion Mass Bounds}

\setcounter{equation}{0}

The numerical solution to the locally approximated Wilson renormalization
group equation can also be used to obtain bounds on the domain of
allowed scalar and fermion masses.  Working in the broken $\gamma_5$-symmetry
phase, we define the $\varphi$ and $\psi$ masses as the zero momentum value
of the one particle irreducible scalar and fermion
two point Schwinger functions, $\Gamma^{(2,0,0)}(p)$ and $\Gamma^{(0,1,1)}(p)$,
respectively, so that
\be
\frac{M_s^2}{\Lambda^2} \equiv \Gamma^{(2,0,0)}(p=0)
\label{4.1}
\ee
\be
\frac{m_f}{\Lambda} \equiv {1\over 4}{\rm tr}~\Gamma^{(0,1,1)}(p=0).
\label{4.2}
\ee
Since $Z=1$ in the local action approximation we are employing, these masses
are also the
location of the poles of the propagators when analytically continued to
Minkowski space.

The short distance contributions to the Schwinger
functions are evaluated by numerically integrating the Wilson renormalization
group equation from the initial cutoff $\Lambda$, where the generalized
potential is specified, down to a lower momentum scale, $\Lambda^* =e^{-t^*}
\Lambda$, below which the dynamics can be accurately described perturbatively.
The contributions to the zero momentum Schwinger
functions from degrees of
freedom with momentum below $\Lambda^*$ (infrared contributions) are then
included as (one-loop) perturbative corrections to the generalized potential
$U(\varphi, \sigma; t^*)$ obtained from the numerical integration.  In this
sense, the Wilson renormalization group equation provides a systematic
procedure of generating the effective Lagrangian valid on one distance scale
from that of a more finely grained distance scale.  Integrating all the way
down to zero momentum ($t \to \infty$), the effective generalized potential
is secured as (cf. Eqs.~(\ref{3.21}- \ref{3.23})) as
\bea
U^{eff}(\varphi,\sigma)
 &\equiv & U(\varphi,\sigma;t^*)+{1\over 2}\int_{0\leq |p|\leq 1}
{d^4p\over (2\pi)^4}~{\rm str~ln~}\left[K_{ij}+{\partial^2
U(\varphi,\sigma;t^*)\over \partial \Omega_j \partial\Omega_i}\right]
\nonumber\\
&=&V^{eff}(\varphi)+\sigma G^{eff}(\varphi) \, ,
\label{4.3}
\eea
where
\bea
V^{eff}(\varphi)&=&V(\varphi;t^*) +{1\over 64\pi^2}
\left[(1-V_{\varphi\varphi}^2(\varphi;t^*))
{\rm ln}~(1+V_{\varphi\varphi}(\varphi;t^*))\right.\nonumber\\
& &\qquad\qquad\qquad\qquad \left. +V_{\varphi\varphi}(\varphi;t^*)
+V_{\varphi\varphi}^2 (\varphi;t^*) {\rm ln}~
V_{\varphi\varphi}(\varphi;t^*)\right]\nonumber\\
 & &-{1\over 16\pi^2}\left[(1-G^4 (\varphi;t^*))~{\rm ln}~
(1+G^2 (\varphi;t^*))
+G^2 (\varphi;t^*) \right.\nonumber \\
& &\quad\quad\qquad\qquad \left. +G^4 (\varphi;t^*)~{\rm ln}~G^2
(\varphi;t^*)\right]
\label{4.4}
\eea
\bea
G^{eff} (\varphi)&=& G(\varphi;t^*)+{1\over 32\pi^2}
\left[G_{\varphi\varphi} (\varphi;t^*)+{2G(\varphi;t^*)
G_\varphi^2 (\varphi;t^*)\over (V_{\varphi\varphi}(\varphi;t^*)
-G^2(\varphi;t^*))}\right] \nonumber \\
& &\qquad\qquad\qquad\qquad \left[1+V_{\varphi\varphi} (\varphi;t^*)
{}~{\rm ln}~{V_{\varphi\varphi}(\varphi;t^*) \over (1+
V_{\varphi\varphi}(\varphi;t^*))}\right]\nonumber\\
 & &-{1\over 16\pi^2}{G(\varphi;t^*)G_\varphi^2(\varphi;t^*) \over
(V_{\varphi\varphi}(\varphi;t^*)
-G^2(\varphi;t^*))} \nonumber\\
 & &\qquad\qquad\qquad \left[1+G^2 (\varphi;t^*)~{\rm ln}~{G^2
(\varphi;t^*)\over
(1+ G^2 (\varphi;t^*))}\right] \, . \nonumber \\
\label{4.5}
\eea

To proceed with the numerical integration of Eq.~(\ref{3.15}), we first
determine the critical value of the mass parameter, $\mu_{crit}^2$, for
all possible choices of the initial coupling constants $\lambda (0)$ and
$g(0)$.  For each particular choice of initial $\lambda (0)$
and $g(0)$ parameters, $\mu_{crit}^2$ is defined so as to separate the
discrete $\gamma_5$-symmetric
phase ($\mu^2(0)>\mu^2_{crit}$) from the spontaneously broken $\gamma_5$-
symmetry phase ($\mu^2(0)<\mu^2_{crit}$).  The broken symmetry ground state
is determined by the location of the zeros of $F(\varphi;t)$.
The value of $\mu^2_{crit}$
is the maximum of $\mu^2(0)$, for a given $\lambda (0)$ and $g(0)$, which
results in a nontrivial zero of $F(\varphi;t)$ as $t$
increases into the infrared.   This is determined
by evaluating $F(\varphi;t)$ for $\mu^2(0)$ well into
the broken phase and then increasing $\mu^2(0)$ until the vacuum value
decreases, rather than increases, as the model is evolved in $t$.  The
transition value of $\mu^2(0)$ delineating the increasing and decreasing
evolution of the vacuum value
is $\mu^2_{crit}$.  Performing this analysis, one also finds that the
model has only a trivial (Gaussian) infrared fixed point in accord with
the conclusions based on lattice simulations.$~^{[12]}$

Once the critical line is determined, a choice of initial parameters
$\lambda(0),\\
g(0)$ and $\mu^2(0)<\mu^2_{crit}$ is made and the $t$
dependence of the generalized
potential terms $F(\varphi;t)$ and $G(\varphi;t)$ is extracted.
This nonperturbative evolution is continued until the the scale $t^*$ is
reached at which point the dynamics of the
system can be accurately described by perturbation theory and the generalized
potential is given via Eqs.~(\ref{3.21}-\ref{3.23}).  The scalar and
fermion masses are then extracted using the (one-loop) perturbative expressions
for the corresponding zero momentum Schwinger functions.  Thus the scalar mass
is given by the
curvature of the effective potential at its minimum via
\bea
\frac{M^2_s}{\Lambda^2}&=&\Gamma^{(2,0,0)}(p=0)
\nonumber\\
 &=& V^{eff}_{\varphi\varphi}(\frac{v}{\Lambda})
\label{4.6}
\eea
and the fermion mass is simply the effective Yukawa coupling constant also
evaluated at the minimum of the effective potential:
\bea
\frac{m_f}{\Lambda} &=& {1\over 4} {\rm tr~}\Gamma^{(0,1,1)}(p=0)\nonumber\\
 &=& G^{eff}(\frac{v}{\Lambda}) \, .
\label{4.7}
\eea

The vacuum expectation value, $v/\Lambda$, is given, through one loop,
by the location of the minimum of the effective potential as
\be
V^{eff}_\varphi (\frac{v}{\Lambda}) = 0 \, .
\label{4.8}
\ee
Letting $v_0$ denote the vacuum expectation value of the the tree level
potential $V(\varphi;t^*)$, then the one
loop shift in expectation value, $\delta v$, is just
\bea
\frac{\delta v}{\Lambda} &=& \frac{v}{\Lambda} -\frac{v_0}{\Lambda} \nonumber\\
 &=& \left\{-{1 \over 32\pi^2}\frac{V_{\varphi\varphi\varphi} }
{V_{\varphi\varphi} }\left[1+V_{\varphi\varphi} ~{\rm ln}~\frac
{V_{\varphi\varphi} }{(1+V_{\varphi\varphi})}\right]\right.
\nonumber\\
 & &\left.\left. \qquad +{1 \over 4\pi^2}\frac{G
G_\varphi }{V_{\varphi\varphi} }
\left[1+G^2 {\rm ln}~\frac{G^2}{(1+G^2 )}\right]\right\}\right|_{\varphi=v_0 /
\Lambda,~t=t^*}~~.
\label{4.9}
\eea

Using these results in Eqs.~(\ref{4.4}-\ref{4.7}), the scalar and fermion
masses are
secured in terms of the
effective potential and effective generalized Yukawa coupling evaluated at
$t^*$ and the tree vacuum value $v_0$ as:
\bea
\frac{M_s^2}{\Lambda^2} &=&\left\{V_{\varphi\varphi}
+{1 \over 32\pi^2}
\left[V_{\varphi\varphi\varphi\varphi}
-\frac{V_{\varphi\varphi\varphi}^2 }
{V_{\varphi\varphi}
(1+V_{\varphi\varphi} )} \right. \right. \nonumber \\
& &\left. \left. \qquad\qquad\qquad\qquad +V_{\varphi\varphi}
V_{\varphi\varphi\varphi\varphi}
{\rm ln}~~\frac{V_{\varphi\varphi} }
{(1+V_{\varphi\varphi} )}\right]\right.
\nonumber\\
 & &\left. -{1 \over 4\pi^2}\left[(G
G_\varphi )_\varphi +\frac{2G^2
G_\varphi^2 }{(1+G^2 )}-\frac{V_{\varphi\varphi\varphi} }
{V_{\varphi\varphi} }G G_\varphi \right. \right.
\nonumber\\
 & &\left. \left. \qquad\qquad +\left(3G_\varphi^2  +G
G_{\varphi\varphi}
-\frac{V_{\varphi\varphi\varphi} }
{V_{\varphi\varphi} }
G G_\varphi  \right)\right. \right.
\nonumber \\
& &\left. \left. \left. \qquad\qquad\qquad G^2 ~{\rm ln}~\frac{G^2}
{(1+G^2) }\right]\right\}\right|_{\varphi=v_0
/\Lambda,~t=t^*}
\label{4.10}
\eea
\bea
\frac{m_f}{\Lambda}&=&\left\{ G -{1 \over 32\pi^2}
\left[\frac{G_\varphi
V_{\varphi\varphi\varphi} }
{V_{\varphi\varphi} }+G_{\varphi\varphi}+\frac{2 G
 G_\varphi^2 }{(V_{\varphi\varphi}-G^2 )}\right]\right. \nonumber\\
& &\left. \qquad\qquad\qquad\qquad\qquad \left[1+V_{\varphi\varphi}
{}~{\rm ln}~\frac{V_{\varphi\varphi} }{(1+
V_{\varphi\varphi} )}\right]\right.
\nonumber\\
 & &\left. +{1\over 4\pi^2}\left[\frac{G
G_\varphi^2 }{V_{\varphi\varphi}}
+{1 \over 4}\frac{G G_\varphi^2}{(V_{\varphi\varphi}
-G^2 )}\right]\right. \nonumber \\
& &\left. \left. \qquad\qquad\qquad \left[1+G^2 ~{\rm ln}~
\frac{G^2}{(1+G^2 )}\right]\right\}\right|_{\varphi=v_0 /
\Lambda,~t=t^*} \, .
\label{4.11}
\eea
The value of $t^*$ is
ascertained by the desired numerical accuracy of these masses determinations.
As the system evolves into the perturbative regime, the mass values obtained
remain stable as $t^*$ is varied.  We choose for $t^*$ that particular value
such that the computed mass is stable to within an error the size of the
$t$-grid spacing: $10^{-5} < \delta t < 10^{-4}$.

Using the above procedure, the ratios $M_s/\Lambda$, $m_f/\Lambda$ and
$v/\Lambda$ are determined as the initial couplings $\lambda(0)$ and $g(0)$
are varied.  Taking appropriate products of these ratios, we plot in Fig. 6
the allowed $M_s/v$ values as a function of $\Lambda/M_s$ for various
different choices of $\lambda(0)$ and $g(0)$.  Note that the plotted
data is further restricted to satisfy the physically imposed constraint that
$\Lambda/M_s \geq 1$.  This follows since no masses are allowed to become
larger than
the initial UV cutoff $\Lambda$.  An upper bound on the $M_s/v$ values
as a function of $\Lambda/v$ is secured by noting that a dynamical envelope is
formed as the initial $\lambda(0)$ and $g(0)$ couplings are varied.   That is,
for fixed $\Lambda / v$, the different $M_s / v$ values obtained appear to
converge to an upper limit, at least for initial Yukawa couplings $\leq 10$.
\footnote{Due to limitations in the presently used numerical integration
technique, we do not establish the dynamical envelope for larger values of the
initial Yukawa coupling $g(0)$.}

\begin{figure}
\vspace{5.0in}
\caption{$M_s/v$ as a function of $\Lambda/M_s$ for different initial
$\lambda(0)$ and $g(0)$ values.}
\end{figure}

Focusing on the dependence of $M_s/v$ on the initial Yukawa coupling, we note
that for $\lambda(0)=5$, a change in $g(0)$ from 0 to 7 leads to roughly a
15-20\% increase in $M_s/v$.  On the other hand, for the larger initial value
of $\lambda(0)=15$, the same change in $g(0)$ from 0 to 7 results in only an
6-10\% increase in $M_s/v$.  Thus the effect of the initial Yukawa coupling on
$M_s/v$ diminishes with increasing initial $\lambda(0)$.

For fixed $\lambda(0)$ and $g(0)$, note that all the curves in
Fig. 6 except the one denoted with the crosses ($\times$), $\Lambda/v$
increases as $\Lambda/M_s$ increases.  The contrary situation persists for the
curve with $\lambda(0)=0$, $g(0)=7$ where the maximum scalar mass also
corresponds to the largest $\Lambda/v$.  Furthermore, the curve corresponding
to $\lambda(0)=5$ and $g(0)=7$ has the behavior that $M_s/v$ initially
increases with  $\Lambda/v$ before reaching its maximum and then subsequently
deceases.  Except for these two curves, which are characterized by having
$g(0)>\lambda(0)$, all the other curves exhibit a maximum scalar mass when
$\Lambda/v$ ratio is at its minimum.  The reason for the behavior of these
two curves is that the
initially large $g(0)$ value requires a longer evolution time (or
equivalently a larger initial cutoff) to indirectly feed into an increase in
the scalar mass.

In addition, we can extract the allowed domain of scalar and fermion masses
for a fixed value of the ultraviolet cutoff to scalar vacuum expectation value
ratio $\Lambda/v$.  Figure 7 is just such a plot$~^{[13]}$ corresponding to the
specific choice $\Lambda/v =5$.  The allowed values are those lying interior to
the various boundaries displayed.  We now discuss the origin of each of the
boundary restrictions in turn.  The upper portion of the boundary, denoted by
the cross ($\times$) marks, is the scalar mass triviality bound discussed
above.  Note that
the absolute scalar mass upper bound, for this value of $\Lambda/v$, occurs
for the larger fermion masses and is simply given by $M_s=\Lambda=5v$.  Next,
the square markings ($\Box$) correspond to the triviality bounds on $m_f$.
These points
are obtained by finding the largest $g(0)$ values for a given $\lambda(0)$
which produce an envelope in the $m_f/v$ versus $\Lambda/m_f$ plane.  Finally,
by demanding that the scalar potential, $V(\varphi;t)$, be bounded from below
on all scales, we further restrict the allowed masses.  This vacuum stability
requirement, which corresponds to the condition that
$lim_{|\varphi| \rightarrow \infty} F(\varphi;t) > 0$ for all t, is implemented
by setting $\lambda(0)=0$ and allowing $g(0)$ to vary.  This results in the
points on the boundary in Fig. 7 represented by the diamonds ($\Diamond$).
Note that for
larger fermion masses, the fermion mass triviality condition provides a more
stringent bound than that of vacuum stability.  This is not the case for very
large $\Lambda/v$ ratios where the vacuum stability constraint is always more
stringent$~^{[13]}$ than that due to the fermion triviality.

\begin{figure}
\vspace{4.0in}
\caption{Allowed range of $M_s /v$ and $m_f /v$ values for $\Lambda/v = 5$.}
\end{figure}

\newpage

\section{Conclusions}

The Wilson renormalization group equation provides a systematic
nonperturbative tool for
constructing the action at a particular distance scale given the form of the
action at a more finely grained scale.  The resulant action contains operators
of all mass dimension.  In particular, if one defines the form of the action at
some UV cutoff scale $\Lambda$, then for slightly lower scales, the irrelevant
operators (those of mass dimension greater than the dimensionality of
space-time) can play a highly nontrivial role in the ensuing dynamics.

We constructed the Wilson renormalization group equation for a theory
involving a scalar and a fermion and numerically solved the equation in a local
approximation.  The resulting renormalization group trajectories were analyzed
and we established triviality and vacuum stability mass bounds on both the
scalar and fermion degrees of freedom.  Clearly, it is desirable to extend
this study beyond the local approximation to include the anomalous dimensions
(nontrivial wavefunction renormalizations) for the scalar and fermion as well
as higher dimension derivative couplings.  At the present time, the
complications limiting such an extension are purely of a numerical nature and
we plan to address these issues in future studies.  In addition, the analysis
can also be enlarged to allow for the more realistic case of scalars and
chiral fermions transforming nontrivially under an internal non-Abelian Lie
group and also via the inclusion of gauge bosons.  This later step can prove
technically challenging since the hard cutoff we employ here breaks the
manifest gauge invariance.  Finally, the model can also be made supersymmetric
and an analysis similar to that performed here can yield SUSY mass triviality
bounds.

\newpage

\centerline{\large\bf Appendix:}
\centerline{\large\bf Supermatrices, Supertraces and Superdeterminants}

\setcounter{equation}{0}
\renewcommand{\theequation}{A.\arabic{equation}}

When discussing theories containing both bosons and fermions,
it often proves convenient to combine terms possessing different
Grassmann algebra character into a single supermatrix structure.
In this appendix, we review some of the properties of the
supermatrix
\be
  M = \pmatrix{M_{BB} & M_{BF} \cr
        M_{FB} & M_{FF}\cr} \, ,
\label{A.1}
\ee
where the square submatrices $M_{BB}$ and $M_{FF}$ contain
only even elements of a Grassmann algebra while $M_{BF}$ and
$M_{FB}$ contain odd elements of the Grassmann algebra.
In particular, $M_{BB}$ and $M_{FF}$ are so defined such
that the components of $M_{BB}$ are composed of some ordinary
c-numbers (Grassmann number zero) as well as possibly some
higher even elements of the algebra, while the components of
$M_{FF}$ are of Grassmann number two or higher.

We define the supertrace (str) of the supermatrix $M$ via
\be  {\rm str}~M = {\rm tr}~M_{BB} - {\rm tr}~M_{FF} \, ,
\label{A.2}
\ee
where ${\rm tr}$ denotes the ordinary trace.  This definition is
chosen so that the familiar trace property
\be  {\rm str}~(MN) = {\rm str}~(NM)
\label{A.3}
\ee
is still guaranteed to hold.  Here the multiplication law for
supermatrices is defined precisely the same as for ordinary
matrices.  Using the definition of the ${\rm str}$ operation, we
next define the superdeterminant (sdet) by
\be  {\rm sdet}~ M = e^{{\rm str}\,\ln~M} \, .
\label{A.4}
\ee
Using the ${\rm str}$ property (A.3), it is straightforward to
establish that
\be  \delta ({\rm sdet}~M) = {\rm sdet}~M\cdot {\rm str}~(M^{-1}\delta M)
\label{A.5}
\ee
and that
\be {\rm sdet}~(MN) = {\rm sdet}~M \cdot {\rm sdet}~ N \, .
\label{A.6}
\ee

Now consider decomposing the supermatrix $M$ as
\be  M = \pmatrix{M_{BB} & 0 \cr
     M_{FB} & 1\cr} \cdot
         \pmatrix{1 & M_{BB}^{-1} M_{BF} \cr
                  0 & N_{FF} \cr} \, ,
\label{A.7}
\ee
where
\be  N_{FF} = M_{FF} - M_{FB} M_{BB}^{-1} M_{BF} \, .
\label{A.8}
\ee
{}From property (A.6), it follows that
\be  {\rm sdet}~M = {\rm sdet} \pmatrix {M_{BB} & 0 \cr
     M_{FB} & 1 \cr} \cdot {\rm sdet}
                 \pmatrix {1 & M_{BB}^{-1} M_{BF} \cr
           0 & N_{FF}\cr} \, .
\label{A.9}
\ee
Because each of the supermatrices appearing on the right hand
side in the decomposition (\ref{A.7}) has zeroes in either the
upper right or lower left off diagonal blocks, their respective
${\rm sdet}$ can be directly evaluated using the definition (\ref{A.4})
and the ${\rm str}$ property (\ref{A.2}).  For example,
\be
    \begin{array}{rcl}
  {\rm sdet} \pmatrix{M_{BB} & 0 \cr
                                   M_{FB} & 1 \cr}
      &=& \exp \left\{ {\rm str}~\ln \left[ \pmatrix
          {1 & 0 \cr
          0 & 1 \cr} + \pmatrix
        {M_{BB} - 1 & 0 \cr
          M_{FB} & 0 \cr} \right] \right\} \\
    &=& \exp \left\{ {\rm str} \sum_{n=1}^\infty \frac{(-)^{n-1}}{n}
       \pmatrix {M_{BB} - 1 & 0 \cr
                                M_{FB} & 0 \cr} \right\}\\
    &=& \exp \left\{ {\rm tr} \sum_{n=1}^\infty
     \frac{(-)^{n-1}}{n} (M_{BB} - 1)^n \right\} \\
    &=& \exp \left\{ {\rm tr}~\ln M_{BB} \right\} \\
     &=& {\rm det}~ M_{BB} \, ,
   \end{array}
\label{A.10}
\ee
where ${\rm det}$ is the ordinary determinant.  Similarly, one finds
\bea
    {\rm sdet} \pmatrix
      {1 & M_{BB}^{-1} M_{BF} \cr
      0 & N_{FF} \cr} &=& \exp \left\{ (-) {\rm tr}~\ln
       N_{FF} \right\}  \nonumber \\
     &=& ({\rm det}~N_{FF})^{-1} \, .
\label{A.11}
\eea
Here the negative sign appearing in the exponential, which in
turn leads to the inverse determinant, is a direct consequence
of the supertrace definition, Eq.~(\ref{A.2}).  Combining terms, we
secure
\be  {\rm sdet}~M = \frac{{\rm det}~M_{BB}}{{\rm det}~N_{FF}} \, .
\label{A.12}
\ee
Alternately, we could have chosen to decompose $M$ as
\be  M = \pmatrix
       {N_{BB} & M_{BF} M_{FF}^{-1} \cr
         0 & 1 \cr} \cdot
     \pmatrix {1 & 0 \cr
                        M_{FB} & M_{FF} \cr} \, ,
\label{A.13}
\ee
with
\be  N_{BB} = M_{BB} - M_{BF} M_{FF}^{-1} M_{FB} \, .
\label{A.14}
\ee
So doing, one finds
\be  {\rm det}~M = \frac{{\rm det}~N_{BB}}{{\rm det}~M_{FF}} \, .
\label{A.15}
\ee

\bigskip

B.H. and S.T.L. thank the Fermilab theory group for their hospitality during
the course of this investigation. This work was supported in part by the
U.S. Department of Energy under contract DE-AC02-76ER01428 (Task B).

\end{document}